\documentclass[preprintnumbers,amsmath,amssymb,floatfix,11pt,prd,onecolumn,superscriptaddress,nofootinbib]{revtex4}
\usepackage[utf8]{inputenc}
\usepackage{diagbox}
\usepackage{latexsym}
\usepackage{epsfig}
\usepackage{epstopdf}
\usepackage{float}
\usepackage{graphicx}
\usepackage{amssymb}
\usepackage{amsmath}
\usepackage{dcolumn}
\usepackage{subfigure}
\usepackage[caption=false]{subfig}
\usepackage{bm}
\usepackage{color}
\usepackage{comment}
\usepackage[shortlabels]{enumitem}
\usepackage{subcaption}
\usepackage{multirow}
\usepackage{xcolor}
\begin{document}
\title{\bf Magnetic Reconnection and Energy Extraction from a Rotating Black Hole in a Four-dimensional Einstein-Gauss-Bonnet Gravity}

\author{Abdul Malik Sultan}
\email{ams@uo.edu.pk, maliksultan23@gmail.com}
\affiliation{Department of Mathematics, University of Okara, Okara-56300, Pakistan}

\author{Muhammad Israr Aslam}
\email{mrisraraslam@gmail.com, israr.aslam@umt.edu.pk}
\affiliation{Department of Mathematics, School of Science, University of Management and Technology, Lahore-54770, Pakistan}

\author{Muhammad Nawaz}
\email{mnawazzz158@gmail.com}
\affiliation{Department of Mathematics, University of Okara, Okara-56300, Pakistan}

\author{Rubab Manzoor}
\email{rubab.manzoor@umt.edu.pk}
\affiliation{Department of Mathematics, School of Science, University of Management and Technology, Lahore-54770, Pakistan}

\author{Ke Wang}
\email{kkwwang2025@163.com}
\affiliation{School of Material Science and Engineering, Chongqing Jiaotong University, Chongqing 400074, China}

\author{Hamood Ur Rehman}
\email{hamood@uo.edu.pk}
\affiliation{Department of Mathematics, University of Okara, Okara-56300 Pakistan}

\author{Yakup Yildirim}
\email{yyildirim@biruni.edu.tr}
\affiliation{Department of Computer Engineering, Biruni University, Istanbul–34010, Turkey.}

\begin{abstract}
Recently, Comisso and Asenjo proposed a new mechanism for energy extraction based on magnetic reconnection of plasma within the ergosphere. In this paper, we analyze the power and efficiency of energy extraction through magnetic reconnection in a rotating four-dimensional Einstein-Gauss-Bonnet gravity black hole. Firstly, we analyze the background properties of this spacetime and its physical quantities, including the event horizon, the boundary of the ergosphere, and the size of the ergosphere. We analyze the magnetic reconnection in circular orbits and investigate the energy extraction region, as well as the power and efficiency of energy extraction. Our results indicate that energy extraction remains feasible even for a low spin parameter of $0.4$, which is well below the previously reported threshold. We also find that the energy extraction power exceeds that of the Blandford-Znajek mechanism. The Gauss-Bonnet coupling parameters $\alpha$ lower the spin threshold for energy extraction. Similarly, we analyze the energy extraction region in the plunging regime, together with the corresponding power and efficiency. We find that energy extraction is possible even for a low spin parameter as $0.22$. We also observe that the Gauss-Bonnet coupling parameter $\alpha$ further lowers the spin threshold. This behavior is consistent with that observed in the circular orbit case. Finally, we compare the energy extraction power in the plunging and circular orbit regimes and find that the plunging regime yields a higher energy extraction power than the circular orbit case.
\end{abstract}
\maketitle

\section{INTRODUCTION }
Black holes (BHs) are among the most fascinating and extreme objects in the Universe. Their existence was first predicted by Albert Einstein's theory of general relativity (GR), which describes gravity as the curvature of spacetime. Owing to their profound gravitational influence, BHs play a crucial role in a wide range of astrophysical phenomena. Over the past few decades, numerous astronomical observations and discoveries have provided compelling evidence for the existence of BHs. Among the most significant milestones are the direct detection of gravitational waves \cite{abbott2016observation,abbott2016gw151226,abbott2017gw170104,abbott2017gw170814} from BH mergers and the groundbreaking images released by the Event Horizon Telescope (EHT) \cite{akiyama2019first,event2022first} collaboration. These remarkable achievements have transformed BH from purely theoretical predictions into observationally established astrophysical objects. Consequently, BH have become central subjects in modern theoretical and observational physics, attracting considerable attention from the scientific community. Their rich physical properties and profound implications for gravity, astrophysics, and fundamental physics continue to motivate extensive research and exploration. In this perspective, BHs arise as a natural consequence of Einstein’s theory of GR \cite{penrose1965gravitational}. In modern theoretical physics, they also play a significant role in ongoing attempts to establish a consistent framework that unifies GR with quantum mechanics. The no-hair theorem is a fundamental result within the framework of GR, stating that asymptotically flat, stationary, and axisymmetric BH solutions of the Einstein field equations can be completely characterized by only three parameters, such as the mass $M$, the electric charge $Q$, and the angular momentum $j$ \cite{carter1971axisymmetric,newman1965metric}.


\par Among the various alternatives and extensions to GR, higher-curvature theories of gravity have emerged as theoretically well-motivated frameworks. Such theories naturally incorporate corrections to Einstein's gravity at high energy scales and are often regarded as promising candidates for exploring gravitational phenomena beyond the classical regime. These theories replace the linear relation between energy-momentum and spacetime curvature in GR with more general functional forms that include higher-order curvature terms. Such modifications are often treated with caution due to Lovelock’s theorem \cite{lovelock1971einstein}, as it is generally assumed that higher-order curvature contributions vanish in four-dimensional spacetime or $D\le4$. This apparent restriction limits higher-curvature gravity theories to dimensions $D>4$, where they arise naturally in string theory and other approaches to quantum gravity. In four-dimensional spacetime, the Lagrangian of the Einstein-Gauss-Bonnet (EGB) theory is topologically invariant. Thus, to consider the dynamical impact of EGB gravity, one is generically
needed to work in higher dimensions \cite{eng2}. The simplest nontrivial extension of the Einstein-Hilbert action is provided by the Gauss-Bonnet (GB) term, which is quadratic in the curvature tensors and is defined as
\begin{equation}
S^{\mathrm{GB}}_{D}= \alpha \int d^{D}x \,\sqrt{-g}\,\Big( R^{abcd}R_{abcd} - 4R^{ab}R_{ab} + R^{2} \Big)
\equiv \alpha \int d^{D}x \,\sqrt{-g}\,\mathcal{G},
\label{first}
\end{equation}
where $\alpha$ denotes the GB coupling constant and $\mathcal{G}$ is the GB invariant. In four spacetime dimensions $D=4$, the GB term becomes a total derivative and therefore does not contribute to the gravitational field equations.
 
\par Recently, Glavan and Lin \cite{glavan2020einstein} showed that the EGB term can become dynamical in four-dimensions through a suitable rescaling of the coupling constant $\alpha$, such as $\alpha \rightarrow\frac{\alpha}{D-4}$, and taking the limit $D\rightarrow4$. This framework has been widely used to construct a variety of solutions, including BHs with thermodynamic and geometrical analyses \cite{ghosh2020generating,konoplya2020stability,singh2020thermodynamics,wei2020extended}, electrically charged configurations \cite{fernandes2020charged,zhang2020superradiance}, and spacetimes sourced by magnetic charge or nonlinear electrodynamics \cite{jusufi2020nonlinear,abdujabbarov2020dynamics}. Within this setting, phenomena such as light deflection by BHs have been revisited \cite{islam2020gravitational,jin2020strong}, along with studies of quasinormal modes \cite{churilova2021quasinormal,mishra2020quasinormal}, and optical appearance of BH shadows \cite{zeng2023optical,zubair20234d,konoplya2020quasinormal,zeng2020shadows,rayimbaev2022shadow,zeng2026shadows}. Exotic compact objects, including Morris-Thorne type traversable wormholes and thin-shell constructions, have also been investigated in this context \cite{jusufi2020wormholes,chakraborty2025traversable}.
\par The idea of such dimensional reduction sparked a lot of debate about the consistency of a four-dimensional formulation, with many concerns raised in the literature \cite{gurses2020there,shu2020vacua,ai2020note}. To overcome this limitation, two separate groups independently addressed these foundational questions \cite{hennigar2020taking,fernandes2020derivation}. They adopted the same rescaling procedure originally proposed by Glavan and Lin to construct consistent formulations of what is now referred to as four-dimensional EGB gravity. They modified the gravitational equations by adding a scalar field to the action, thus preserving Lovelock's theorem and rendering the four-dimensional EGB a member of the Horndeski family. The first one uses conformal rescaling methods similar to the one used to obtain the $ D \to 2$ dimensional limit of GR \cite{mann1993d}, while the second one uses the Kaluza-Klein dimensional reduction procedure \cite{lu2020horndeski}. Even though these approaches produce equivalent theories (up to trivial field redefinitions), there is an important distinction: the Kaluza-Klein framework introduces additional terms in the gravitational field equations that depend on the curvature of the maximally symmetric ($D-4$)-dimensional space. When these additional contributions vanish, one arrives at the four-dimensional EGB action contribution as:
\begin{equation}
S_{\text{4D}}^{\text{GB}} = \alpha \int d^4 x \sqrt{-g} \left[ \phi \mathcal{G} + 4 G_{ab} \nabla^a \phi \nabla^b \phi - 4 (\nabla \phi)^2 \Box \phi + 2 (\nabla \phi)^4 \right], 
\end{equation}\label{third}
where the scalar field gradients are defined through \( (\nabla \phi)^2 \equiv g^{ab} \nabla_a \phi \nabla_b \phi \) and \( (\nabla \phi)^4 \equiv \left[ (\nabla \phi)^2 \right]^2 \), with \( \phi \) representing the newly introduced scalar degree of freedom. This contribution supplements the standard Einstein-Hilbert term in the complete theory, effectively modifying GR. What's remarkable is that the static, spherically symmetric BH solutions emerging from the resulting field equations are identical to those obtained from the naive \( D \to 4 \) limit of \( D>4 \) solutions presented in \cite{glavan2020einstein}, yet without ever invoking a higher-dimensional spacetime. Subsequent works have shown that $4$D EGB gravity is a phenomenologically viable alternative to Einstein's theory \cite{charmousis2022astrophysical,zanoletti2024cosmological}, but the physical interpretation and observational consequences of these higher-curvature corrections are still under investigation \cite{fernandes20224d}. Compact astrophysical objects, especially neutron stars, are the best place to test GR and modified gravity theories. Any viable alternative theory should be able to explain the observational properties of known compact objects and predict the gravitational wave signals from sources in the so-called mass gap, the theoretically mysterious regime between the maximum mass of neutron stars and the minimum mass of astrophysical BH. These tight observational constraints are one of the best ways to test four-dimensional EGB gravity against observations.
\par After the foundational work on relativistic stars \cite{doneva2021relativistic}, many studies have been done on neutron stars with realistic equations of state and found that positive GB coupling allows for larger maximum masses that approach the BH mass limit, closing the mass gap \cite{saavedra2025neutron}. Anisotropic neutron star solutions exhibit modified mass-radius relations compared to isotropic configurations \cite{singh2022anisotropic}. Beyond neutron stars, the framework has been applied to quark stars with unified interacting equations of state \cite{gammon2024quark}, strange quark stars \cite{banerjee2021strange}, electrically charged quark stars \cite{pretel2022electrically}, color-flavor locked strange stars \cite{banerjee2021color}, and white dwarfs showing deviations from the Chandrasekhar limit \cite{pretel2025white}. Complementary studies have explored barotropic equations of state in Einstein-Maxwell-GB gravity \cite{hansraj2025barotropic} and gravitational collapse dynamics \cite{jaryal2023spherical}. These investigations demonstrate that four-dimensional EGB gravity yields rich phenomenology with testable predictions distinguishing it from GR.
\par Another remarkable prediction of GR is that a rotating BH possesses extractable energy that can, in principle, be extracted through appropriate physical processes. In \cite{christodoulou1970reversible}, authors first demonstrated that a Kerr BH with mass $M$, and spin parameter $a$, a fraction of the BH mass equal to $M_{irr}=M\sqrt{\frac{1}{2}(1+\sqrt{1-a^2})}$ is irreducible. So, the maximum energy that can be extricated from a BH without violating the second law of thermodynamics corresponds to the rotational energy $E_{rot}=(1-\sqrt{\frac{1}{2}(1+\sqrt{1-a^2})})Mc^2$. Substituting $a=1$, the maximum extracted energy can be turns out to be $E_{\rm rot} \approx 0.29\, M c^2$. There is no conflict between the second law of BH thermodynamics and energy extraction from rotating BHs. 
\par The idea of extracting rotational energy from a BH can be traced back to the pioneering work of Penrose \cite{penrose1969gravitational,penrose1971extraction}, explained it by designing a thought experiment which describe a particle fission $0\rightarrow1+2$ occur in the ergosphere of a Kerr BH. The essence of the Penrose process relies on conservation law. In practice, if the angular momentum of particle $1$ is opposite to the BH rotation, the energy of particle $1$, from the infinity observer’s viewpoint, is negative, meaning that to satisfy conservation laws, the energy of particle $2$, which escaped to infinity should be larger than that of the initial particle $0$. However, there are some limitations to the Penrose process that make it an inefficient energy extraction method in astrophysical scenarios. More precisely, the Penrose process is not expected to lead to the extraction of significant rotational energy from astrophysical BHs to serve as an explanation of high energy astrophysics phenomena \cite{wald1974energy}. However, alternative mechanisms were proposed for extracting BH rotational energy, such as collisional Penrose process \cite{piran1975high}, the magnetohydrodynamic Penrose process \cite{takahashi1990magnetohydrodynamic}, superradiant scattering \cite{teukolsky1974perturbations} and the Blandford-Znajek (BZ) process \cite{blandford1977electromagnetic}.
\par In addition to the traditional energy extraction mechanisms, recent studies \cite{koide2008energy,comisso2021magnetic} have proposed magnetic reconnection as a novel process for extracting energy from the ergosphere of rotating BHs, with potentially higher efficiency than earlier methods. Magnetic reconnection occurs in the plasma rotating around a BH in the equarorial plane. The frame dragging effect of a rapidly rotating BH generates antiparallel magnetic field configurations in the equatorial plane. This magnetic field configuration gives rise to an equatorial current sheet, which serves as the site of fast plasmoid-mediated magnetic reconnection. The process converts magnetic energy into plasma kinetic energy and generates two oppositely directed plasma outflows within the reconnection layer \cite{comisso2021magnetic}. One of these plasma streams falls into the event horizon on a retrograde trajectory with highly negative energy, while the other gains positive energy and escapes to infinity along a prograde orbit. As a result, energy can be extracted from the rotating BH. Energy extraction through magnetic reconnection depends on local magnetic fields and antiparallel magnetic field structures, without requiring the large-scale magnetic field background needed in the magnetic Penrose and BZ processes. This makes the mechanism more likely to operate in the presence of strong magnetic-field loops within the BH ergosphere. Comisso and Asenjo showed that, in kerr BH, the maximum efficiency of energy extraction through magnetic reconnection can reach $3/2$ in the regime of high plasma magnetization and strong magnetic effects \cite{comisso2021magnetic}. In some cases, the power extracted through magnetic reconnection can exceed that of the BZ process, making magnetic reconnection an efficient mechanism for extracting energy from BHs. Consequently, investigations of energy extraction via magnetic reconnection have been extended to a wide range of rotating BHs \cite{wei2022effects,israr1,khodadi2022magnetic,carleo2022energy,wang2022extracting,li2023energy,l2023energy,zhang2024energy,zhang2024,khodad2023harvesting,shaymatov2024kerr,rodriguez2025energy,long2025magnetic,zeng2025energy,wang2025energy,zeng2025e2,eshtursunov2026energy,eshtursunov2026magnetic,yao2026energy,enhanced,yuchih2025energy,cheng2025extractin,alipour2026repetitive}, providing valuable insights for testing GR as well as modified theories of gravity.
\par Inspired by these studies, we explore energy extraction through magnetic reconnection  of a rotating BH in the framework of four-dimensional EGB gravity \cite{kumar2020rotating}. Our aim is to investigate how the spin parameter $a$ and the GB coupling parameter $\alpha$ influence the magnetic reconnection process in the rotating four-dimensional EGB BH and its associated energy extraction power and efficiency. In particular, we examine whether this BH model provides a more efficient mechanism for energy extraction than previously studied rotating BHs spacetimes. We further demonstrate that magnetic reconnection driven energy extraction remains feasible even at low spin, indicating an improvement in the efficiency of the process. This enhancement arises from the effect of the GB coupling, which modifies the spacetime geometry and effectively lowers the critical spin threshold required for efficient energy extraction.
\par This paper is organized as, In section {\bf II}, we briefly define the background of the four-dimensional EGB BH model, along with a discussion of its geometrical and physical properties. In section {\bf III. A}, we describe the magnetic reconnection process in circular orbits. In section {\bf III. B}, we analyzes the allowed parameter space for energy extraction. While section {\bf III. C} is devoted to discussing the corresponding power and efficiency of energy extraction. In section {\bf IV. A}, we present the magnetic reconnection mechanism in the plunging region, while in section {\bf IV. B}, we analyze the associated power of energy extraction. Finally, in last section, we put our conclusions and discussions.

\section{A Brief Review Of The Rotating Black Hole in four-dimensional EGB Gravity} 
\begin{figure}[H]
\begin{center}
\subfigure[~$\alpha=0.001$]{\includegraphics[width=4.7cm,height=4.3cm]{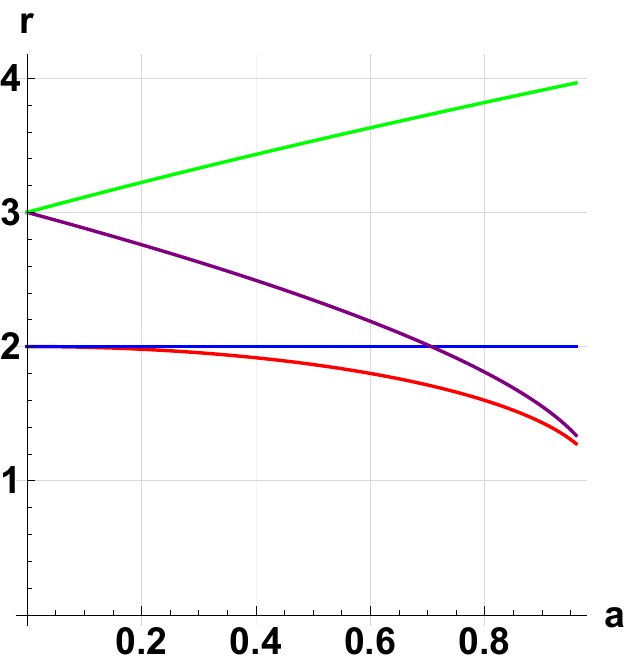}}
\subfigure[~$\alpha=0.01$]{\includegraphics[width=4.7cm,height=4.3cm]{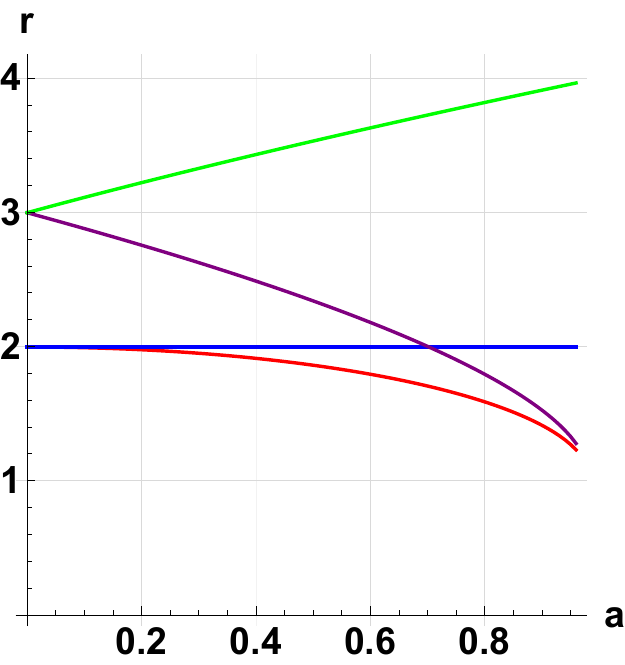}}
\subfigure[~$\alpha=0.1$]{\includegraphics[width=4.7cm,height=4.3cm]{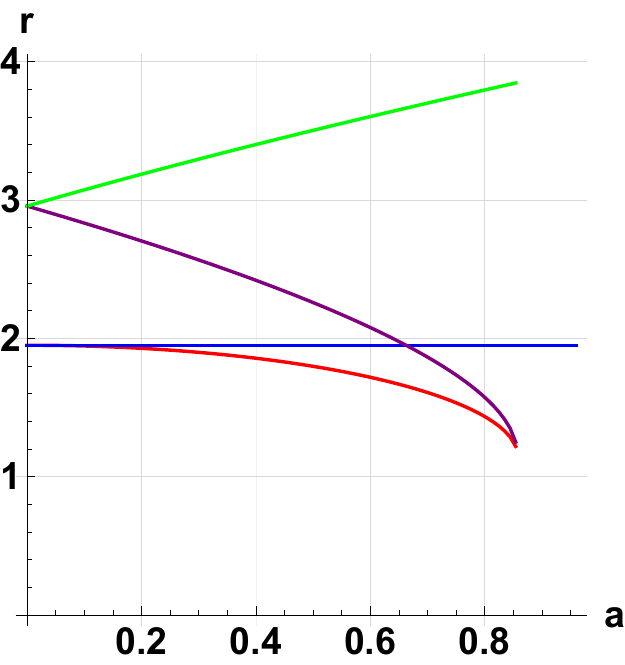}}
\subfigure[~$\alpha=0.2$]{\includegraphics[width=4.7cm,height=4.3cm]{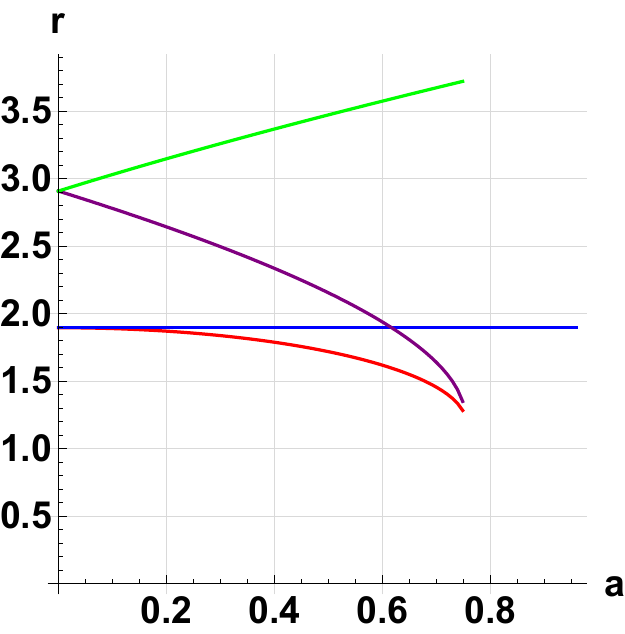}}
\subfigure[~$\alpha=0.3$]{\includegraphics[width=4.7cm,height=4.3cm]{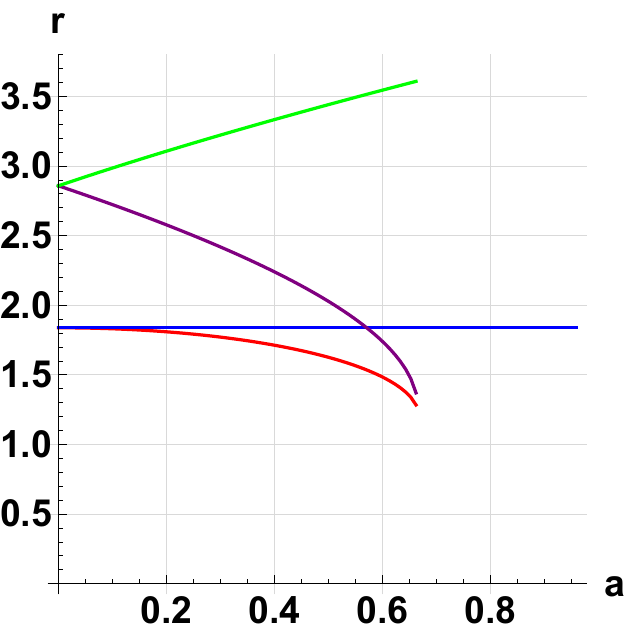}}
\subfigure[~$\alpha=0.4$]{\includegraphics[width=4.7cm,height=4.3cm]{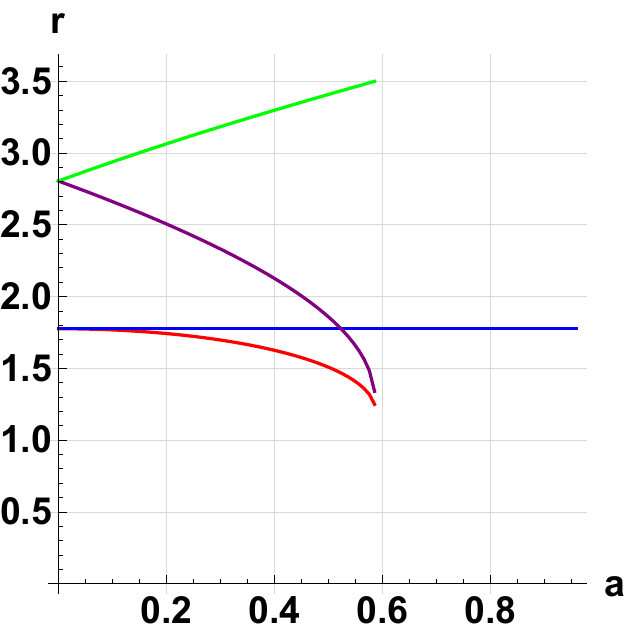}}
\caption{Plots showing the photon sphere radius, ergosphere boundary and event horizon with respect to $a$, under different values of $\alpha$. The green, purple, blue and red curves correspond to the retrograde orbits, prograde orbits, the boundary of the ergosphere, and the event horizon, respectively.}
\label{fig1}
\end{center}
\end{figure}

The action corresponding to EGB gravity in \(D\)-dimensional spacetime can be expressed as follows \cite{kumar2020rotating,heydari2021thin,israr2026visual}
\begin{equation} 
S_{\mathrm{EGB}}=\int d^{D}x\,\left(L_{EH} +\alpha L_{GB}\right)\sqrt{-g},\label{EGBAction}
\end{equation} 
where 
\begin{equation} L_{EH}=R,\qquad L_{GB}=R^{\beta\gamma\tau\chi}-R_{\beta\gamma\tau\chi}-4R^{\beta\gamma}R_{\beta\gamma}+R^{2},
\label{EGBComponents}
\end{equation}
here, \(g\) denotes the determinant of the metric tensor. Furthermore, \(R\), \(R_{\beta\gamma}\), and \(R_{\beta\gamma\tau\chi}\) represent the Ricci scalar, Ricci tensor, and Riemann curvature tensor associated with the spacetime geometry. By the variation of the action (\ref{EGBAction}) with respect to the metric tensor \(g_{\beta\gamma}\), one obtains the corresponding gravitational field equations 
\begin{equation}
G_{\beta\gamma}+\alpha H_{\beta\gamma}=0,\label{fieldequation}
\end{equation}
here, \(G_{\beta\gamma}\) represents the Einstein tensor, and \(H_{\beta\gamma}\) is specified by the following relation
\begin{equation}
H_{\beta\gamma}=2\left(RR_{\beta\gamma}-2R_{\beta\tau}R^{\tau}_{\gamma}-2R_{\beta\tau\gamma\chi}R^{\tau\chi}-R_{\beta\tau\chi\psi}R_{\gamma}^{\ \tau\chi\psi}\right)-\frac{1}{2}L_{GB}g_{\beta\gamma}.\label{HBeatafunction}
\end{equation}
It is important to note that deriving rotating BH solutions in four-dimensional EGB gravity directly from the vacuum field equations is highly nontrivial. At present, no exact analytical solution for rotating configurations is known within this framework. Existing results are largely based on physically motivated constructions that reproduce the essential characteristics of rotating BHs. In this context, Kumar and Ghosh employed the Newman-Janis algorithm on a static solution in four-dimensional EGB gravity to obtain a stationary and axisymmetric rotating BH metric \cite{kumar2020rotating}. Expressed in Boyer-Lindquist (BL) coordinates, the resulting spacetime is given by in the following form \cite{kumar2020rotating,heydari2021thin,israr2026visual}
\begin{equation}
\begin{aligned}
ds^{2} ={}&-\left(\frac{\Delta a^{2}\sin^{2}\theta}{\Sigma}\right)dt^{2}+\frac{\Sigma}{\Delta}dr^{2}-2a\sin^{2}\theta\left(1-\frac{\Delta-a^{2}\sin^{2}\theta}{\Sigma}\right)dt\,d\phi\\[2mm]&+\Sigma d\theta^{2}+\sin^{2}\theta\left[\Sigma+a^{2}\sin^{2}\theta\left(2-\frac{\Delta a^{2}\sin^{2}\theta}{\Sigma}\right)\right]d\phi^{2},
\end{aligned}
\label{EBGMetric}
\end{equation}
along with
\begin{equation}
\Delta=r^{2}+a^{2}+\frac{r^{4}}{2\alpha}\left[1-\sqrt{1+\frac{8\alpha M}{r^{3}}}\right],\qquad \Sigma=r^{2}+a^{2}\cos^{2}\theta.\label{Metricdeltasum} 
\end{equation}
In the limit \(\alpha \to 0\), the metric reduces to the rotating solution of GR, namely the Kerr spacetime. For \(a=0\), it recovers the static, spherically symmetric BH solution in four-dimensional EGB gravity. Furthermore, when both \(\alpha =0\) and \(a =0\), then the metric reduced to the Schwarzschild BH solution. The horizons of the BH are obtained by solving $\Delta=0$. The largest positive root defines the event horizon, $r_{+}$, while the smaller positive root defines the inner horizon,$r_{-}$. The boundary of the ergosphere is determined by setting $g_{tt}=0$; this has one root that lies outside the event horizon. Now, we consider the equation of motion of a particle in the equatorial plane $\theta=\pi/2$. So, the normalization condition corresponding to the geodesic equation is given as 
\begin{equation}
g_{\mu\nu}\frac{dx^{\mu}}{d\tau}\frac{dx^{\nu}}{d\tau}=-\epsilon,
\label{geodesic}
\end{equation}
where $\epsilon=0$ for photons, and $\epsilon=1$ for massive particles. Moreover, $\tau$ is the affine parameter. In this metric energy $E$ and angular momentum $L$ are conserved quantities, which are expressed as  
\begin{equation}
    P_{t}=g_{tt}\frac{dt}{d\tau}+g_{t\phi}\frac{d\phi}{d\tau}=-E,\ P_{\phi}=g_{t\phi}\frac{dt}{d\tau}+g_{\phi\phi}\frac{d\phi}{d\tau}=L,
\label{energyandmomentum}  \end{equation}
where 
\begin{equation}
 \frac{dx^{\mu}}{d\tau}=g^{\mu\nu}P_{\nu}.
\label{forEandL}
\end{equation}
For $\mu=t$ 
\begin{equation}
\frac{dt}{d\tau}=g^{tt}P_{t}+g^{t\phi}P_{\phi}=g^{tt}(-E)+g^{t\phi}L,
\label{formu}
\end{equation}
and for $\mu=\phi$
\begin{equation}
    \frac{d\phi}{d\tau}=g^{t\phi}P_{t}+g^{\phi\phi}P_{\phi}=g^{t\phi}(-E)+g^{\phi\phi}L.
\label{forphi}
\end{equation}
For the equatorial plane $\theta=\pi/2$, which leads to
\begin{equation}
    \frac{d\theta}{d\tau}=0.
\label{forequatorial}
\end{equation}
Substituting (\ref{formu}), (\ref{forphi}) and (\ref{forequatorial}) into (\ref{geodesic}) along with conserved quantities, as defined in (\ref{energyandmomentum}), the corresponding metric components has the following form as
\begin{align}
\left(\frac{dr}{d\lambda}\right)^2
= [(r^{2}+a^{2})E - aL]^2
- \Delta_r\left[(L-aE)^2 + \mathcal{O}\right]=R(r),
\label{radialfunction}
\end{align}
where $\mathcal{O}$ is Carter constant \cite{carter1968global} satisfying $\mathcal{O}=\mathcal{Q}-(L-aE)^2$ and $\mathcal{Q}$ is the Carter's separation constant and $\lambda$ is the Mino time \cite{mino2003perturbative}, related to proper time $\tau$ by $\frac{d\tau}{d\lambda}=\Sigma$. We consider the circular orbital motion of particles confined to the equatorial plane. Such circular orbits exist from spatial infinity down to the photon sphere. The following conditions determine the radius of the photon sphere as
\begin{equation}
R(r)=0,\quad \quad R'(r)=0.
\end{equation}\label{rconditions}  
There are two solutions: one corresponding to prograde orbits, which co-rotate with the BH, and the other corresponding to retrograde orbits, which counter-rotate with the BH.
 In Fig. {\bf \ref{fig1}}, we plot the boundary of the ergosphere, event horizon and photon sphere radius $r$ as a function of $a$.  For this, we represent prograde orbits with purple curves, the boundary of the ergosphere with blue curves, the event horizon with red curves, and retrograde orbits with green curves. The maximum allowed spin for a four-dimensional EGB BH is less than $1$, and the boundary of the ergosphere is smaller than $2$. From both rows of Fig. {\bf \ref{fig1}}, we can see that when the values of $\alpha$ increase, the maximum allowed spin decreases. This indicates that a low BH spin is crucial for efficient energy extraction. From the top row of Fig. {\bf\ref{fig1}}, we can see that the position of the boundary of the ergosphere, event horizon and the photon sphere radius under different values of $\alpha$ (i.e $\alpha=0.001,~0.01,~0.1$) remain the same. But in the second row we can see that the position of the boundary of the ergosphere and event horizon decreases, but the photon sphere radius increases, under different values of $\alpha$, i.e., $\alpha=0.2,~0.3,~0.4$. The only prograde orbits are considered for energy extraction, as suggested by Fig. {\bf\ref{fig1}}. At the photon sphere radius, the energy and angular momentum of the particles approach infinity. Therefore, energy extraction actually occurs between the boundary of the ergosphere and the photon sphere radius. So, the Keplerian angular velocity for particles in the prograde orbits is given  by 
\begin{equation}
\Omega_{K}=\frac{d\phi/d\tau}{dt/d\tau}=\frac{-g_{t\phi,r}+\sqrt{\left(g_{t\phi,r}\right)^{2}-g_{tt,r}\,g_{\phi\phi,r}}}{g_{\phi\phi,r}}.
\label{keplerianvelocity}
\end{equation}
The standard expressions for the conserved energy $E$ and angular momentum $L$ of a particle on a circular equatorial orbit in a stationary axisymmetric spacetime are given as 
\begin{equation}
E=-\frac{g_{tt}+g_{t\phi}\Omega_{K}}{\sqrt{-g_{tt}-2g_{t\phi}\Omega_{K}-g_{\phi\phi}\Omega_{K}^{2}}},\qquad L=\frac{g_{t\phi}+g_{\phi\phi}\Omega_{K}}{\sqrt{-g_{tt}-2g_{t\phi}\Omega_{K}-g_{\phi\phi}\Omega_{K}^{2}}}.
\label{expressionforEandL}
\end{equation}
Particles that satisfies the circular orbit conditions, equations (\ref{keplerianvelocity}) and (\ref{expressionforEandL}) must yield real values. For this satisfaction, the expression under the square root in (\ref{keplerianvelocity}) and (\ref{expressionforEandL}) must be greater than zero.

\section{EXTRACTING ENERGY FROM A Four-dimensional EGB Rotating BLACK HOLE IN THE CIRCULAR ORBIT REGION}

\subsection{Magnetic Reconnection Process in Circular Orbits}
\begin{figure}[H]
\begin{center}
\subfigure[~$a=0.98,~\alpha=0.3,~r=1.4$]{\includegraphics[width=8cm,height=5cm]{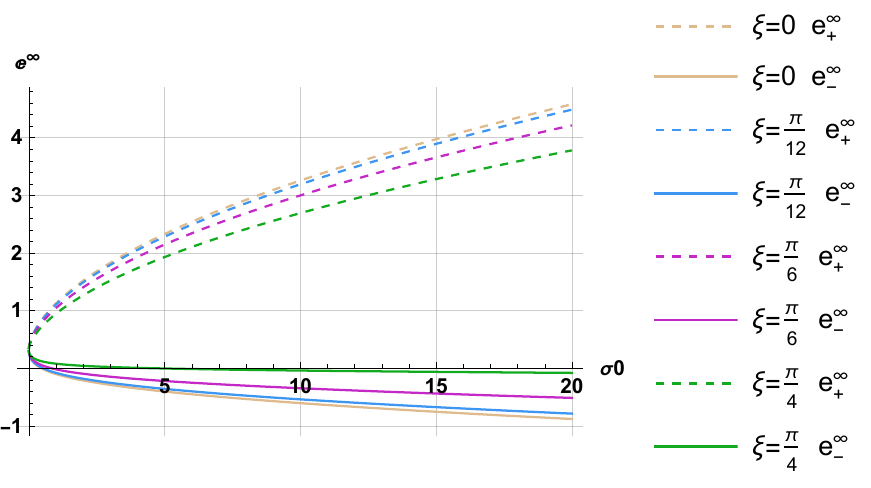}}
\subfigure[~$\alpha=0.06,~r=1.6,~\xi=\pi/12$]{\includegraphics[width=8cm,height=5cm]{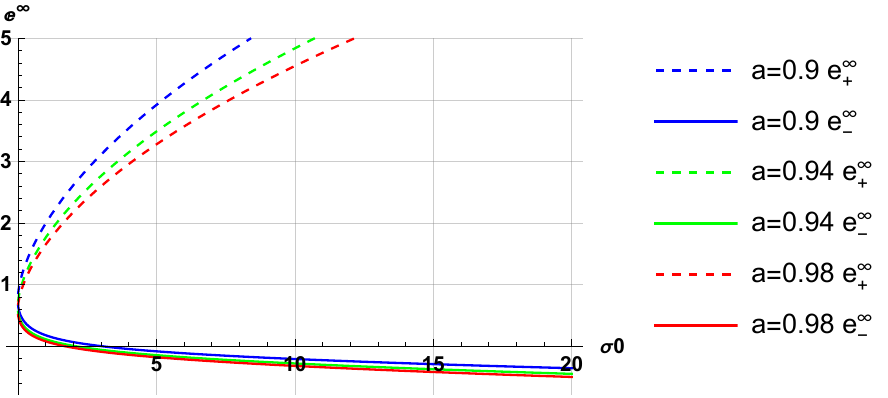}}
\caption{Plots showing the variation of $e^{\infty}_{+}$ and $e^{\infty}_{-}$ with respect to $\sigma_0$. The left panel illustrates the results for different values of $\xi$, whereas the right panel presents the results for different values of the spin parameter $a$.}\label{fig2}
\end{center}
\end{figure}

\begin{figure}[H]
\begin{center}
\subfigure[~$a=0.92,~r=6,~\xi=\pi/12$]{\includegraphics[width=8cm,height=5cm]{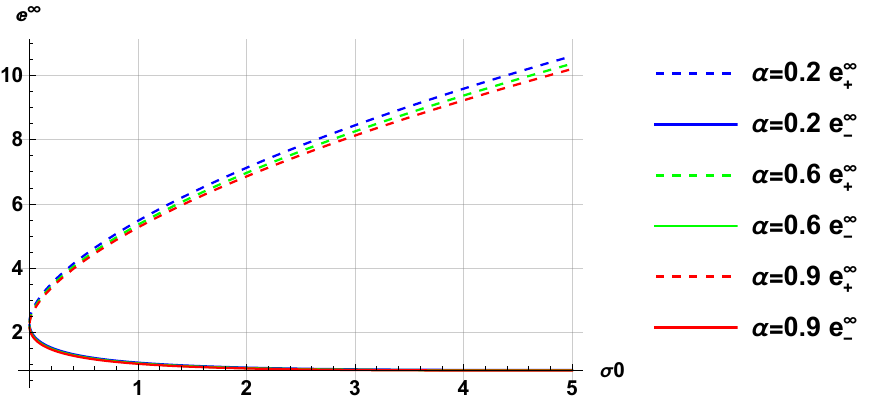}}
\subfigure[~$a=0.92,~r=6.5,~\xi=\pi/12$]{\includegraphics[width=8cm,height=5cm]{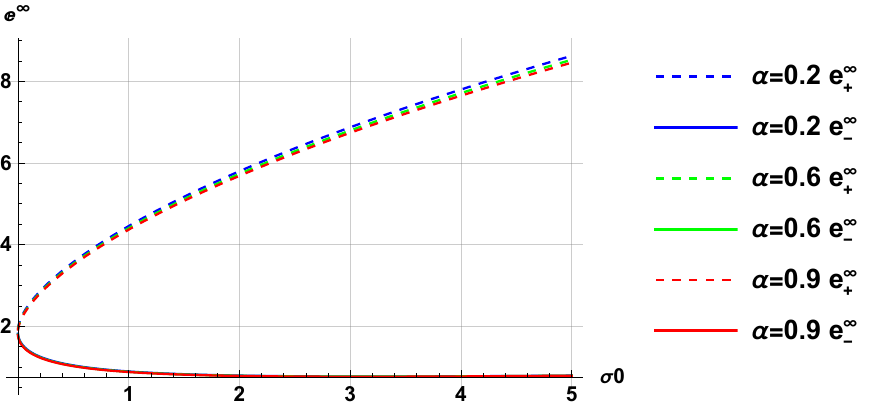}}
\caption{Plots showing the variation of $e^{\infty}_{+}$ and $e^{\infty}_{-}$ with respect to $\sigma_0$ under the different values of $\alpha$.}
\label{fig3}
\end{center}
\end{figure}

In this section, we will analyze the energy extraction via the magnetic reconnection mechanism within the fabric of the four-dimensional EGB BH model. To facilitate the analysis of the plasma density, mass, and energy, a locally non-rotating frame, namely the Zero Angular Momentum Observer (ZAMO) frame, is considered. The advantage of this frame is that the spacetime in this frame is locally Minkowskian for an observer, and this frame also simplifies the corresponding equations and makes their interpretation more transparent. So, the line element in the ZAMO frame has the following form as  
\begin{equation}
ds^{2}=\eta_{\mu\nu}\,d\hat{x}^{\mu}d\hat{x}^{\nu}=-d\hat{t}^{\,2}+\sum_{i=1}^{3}\left(d\hat{x}^{i}\right)^{2}.
\label{lineelement}
\end{equation}
In these two frames, the transformation of coordinates is defined as
\begin{equation}
d\hat{t} = N\,dt,\qquad d\hat{x}^{i}=\sqrt{g_{ii}}\,dx^{i}-N\,\beta^{i}\,dt,
\label{transfor}
\end{equation}
where $N=\sqrt{-g_{tt}+g_{t\phi}^{\,2}/{g_{\phi\phi}}}$ is the lapse function and $\beta^{i}=(0,0,\beta^{\phi})$  is the shift vector. Note that $\beta^{\phi}=\sqrt{g_{\phi\phi}}\omega^{\phi}/N$ with the angular velocity of the frame dragging $\omega^{\phi}=-g_{t\phi}/g_{\phi\phi}$. As discussed above, energy can be extracted from a BH through magnetic reconnection under two conditions:  (i) the formation of a negative energy particle as measured by a distant observer, and (ii) the escape condition for plasma particles accelerated or decelerated by the reconnection process within the ergosphere. The energy-momentum tensor $T^{\mu\nu}$ in one-fluid approximation can be expressed as 
\begin{equation}
T^{\mu\nu}=pg_{\mu\nu}+\omega U^{\mu}U^{\nu}+F^{\mu}_{\delta}F^{\nu\delta}-\frac{1}{4}g^{\mu\nu}F^{\rho\delta}F_{\rho\delta},
 \label{momentumtensor}
\end{equation}
where $p,\omega,U^{\mu}$ and $F^{\mu\nu}$ are the proper plasma pressure, enthalpy density, four-velocity and electromagnetic field tensor, respectively. The energy at infinity density is calculated as
\begin{equation}
e^{\infty}=-Ng_{\mu 0}T^{\mu 0}=N\hat{e}+N\beta^{\phi}\hat {P}^{\phi},
\label{energyatinfinity}
\end{equation}
where the total energy density $\hat{e}$ and the azimuthal component of the momentum density $\hat{P}^{\phi}$ is expressed as
\begin{equation}
\hat{e}=\omega\hat{\gamma}^{2}-p+\frac{1}{2}\left(\hat{B}^{2}+\hat{E}^{2}\right),
\label{edensity}
\end{equation}
\begin{equation}
\hat{P}^{\phi}=\omega\hat{\gamma}^{2}\hat{v}^{\phi}+(\hat{B}\times\hat{E})^{\phi}.
\label{momentumdensity}
\end{equation}
Here $\hat{\gamma}=\hat{U}^{0}=\sqrt{1-\sum_{i=1}^{3}(d\hat{v}^{i})^{2}}$
is the Lorentz factor, and
$\hat{B}^{i}=\epsilon^{ijk}\hat{F}_{jk}/2$
and $\hat{E}^{i}=\eta^{ij}\hat{F}_{j0}=\hat{F}_{i0}\hat{v}^{\phi}$
represent the components of the magnetic and electric fields.
Meanwhile, $\hat{v}^{\phi}$ is the azimuthal component of the outflow velocity of plasma particles for a ZAMO observer. The energy at infinity density $e^{\infty}$ can be divided into the hydrodynamic component $e_{\rm hyd}^{\infty}$ and electromagnetic component $e_{\rm em}^{\infty}$,
\begin{equation}
e^{\infty}=e_{\rm hyd}^{\infty}+ e_{\rm em}^{\infty},
\label{divideinfinitydensity}
\end{equation}
where
\begin{equation}
e_{\rm hyd}^{\infty}=N \hat{e}_{\rm hyd}+N\beta^{\phi}\omega\hat{\gamma}^{2}\hat{v}^{\phi},
\label{ehydrodynamic}
\end{equation}
\begin{equation}
e_{\rm em}^{\infty}=N \hat{e}_{\rm em}+N \beta^{\phi}(\hat{B}\times\hat{E})_{\phi}.
\label{eelectromagnetic}
\end{equation}
Here
$\hat{e}_{\rm hyd}=\omega\hat{\gamma}^{2}-p$
and
$\hat{e}_{\rm em}=(\hat{B}^{2}+\hat{E}^{2})/2$
denote the hydrodynamic and electromagnetic energy densities observed in the ZAMO frame.
Considering an efficient magnetic reconnection process that converts most of the magnetic energy into the kinetic energy of the plasma, the electromagnetic energy at infinity is negligible \cite{comisso2021magnetic}. Further combining with the approximation of incompressible and adiabatic plasma, the energy density at infinity can be well calculated by
\begin{equation}
e^{\infty}=e_{\rm hyd}^{\infty}=N\omega\hat{\gamma}\left(1+\beta^{\phi}\right)-\frac{N p}{\hat{\gamma}}.
\label{newenergydensity}
\end{equation}
In order to investigate the local reconnection process, it is convenient to introduce a local rest frame $x'^{\mu}=(x'^{0},x'^{1},x'^{2},x'^{3})$ for the bulk plasma with Keplerian angular velocity $\Omega_{K}$ in the equatorial plane. In this local rest frame, one can choose the directions such that $x'^{1}$ and $x'^{3}$ are parallel to the radial direction $r$ and azimuthal direction $\phi$, respectively. Then, in the ZAMO frame, the corotating Keplerian velocity can be expressed as \cite{wei2022effects}
\begin{equation}
\hat{v}_{K}=\frac{d\hat{x}^{\phi}}{d\hat{x}^{t}}=\frac{d\hat{x}^{\phi}/d\lambda} {d\hat{x}^{t}/d\lambda}=\frac{\sqrt{g_{\phi\phi}}\,dx^{\phi}/d\lambda-N\beta^{\phi}\,dx^{t}/d\lambda}{N\,dx^{t}/d\lambda}=\frac{\sqrt{g_{\phi\phi}}}{N}\Omega_{K}-\beta^{\phi}.
\label{kepvelocity}
\end{equation}
The coordinate transformation of the vector $\psi$ between the BL coordinates and ZAMO frame is:
\begin{equation}
\hat{\psi}^{0}=N \psi^{0},\qquad\hat{\psi}^{i}=\sqrt{g_{ii}\psi^{i}}-N \beta^{i}\psi^{0},
\label{trans1}
\end{equation}
\begin{equation}
\hat{\psi}_{0}=\frac{\psi_{0}}{N}+\sum_{i=1}^{3}\frac{\beta_{i}}{\sqrt{g_{ii}}}\,\psi_{i},\qquad\hat{\psi}_{i}=\frac{\psi_{i}}{\sqrt{g_{ii}}}.
\label{trans2}
\end{equation}
Finally, employing the relativistic adiabatic incomprehensible ball approach, the hydrodynamic energy at infinity per enthalpy of plasma is \cite{comisso2021magnetic} 
\begin{equation}
e_{\pm}^{\infty}=N \hat{\gamma}_{K}\left[\left(1+\beta^{\phi}\hat{v}_{K}\right)\sqrt{1+\sigma_{0}}\pm\cos\xi\,\left(\beta^{\phi}+\hat{v}_{K}\right)\sqrt{\sigma_{0}}-\frac{\sqrt{1+\sigma_{0}}\mp\cos\xi\,\hat{v}_{K}\sqrt{\sigma_{0}}}{4\hat{\gamma}_{K}^{2}\left(1+\sigma_{0}-\cos^{2}\xi\,\hat{v}_{K}^{2}\sigma_{0}\right)}\right],
\label{epsilon}
\end{equation}
where $\hat{\gamma_{K}}=1/\sqrt{1-\hat{v}^{2}_{K}}$ is the  Lorentz factor corresponding to $\hat{v}_{K}$, $\xi$ is the orientation angle between the azimuthal direction and the magnetic field lines in the equatorial plane, and $\sigma_{0}=B^2_{0}/\omega$ is the plasma magnetization upstream of the reconnection layer. There are two conditions to achieve this energy: (i) decelerated plasma has negative energy, and (ii) accelerated plasma has positive energy for the distant observer \cite{comisso2021magnetic},
\begin{equation}
e_{-}^{\infty}<0,\qquad\Delta e_{+}^{\infty}=e_{+}^{\infty}-\left(1-\frac{\Gamma}{\Gamma-1}\frac{p}{w}\right)=e_{+}^{\infty}>0,
\label{energyconditions}
\end{equation}
here $\Gamma$ is the polytropic index for a relativistic hot plasma, which has a value equal to $4/3$. In Figs. {\bf \ref{fig2}} and {\bf\ref{fig3}}, we plot the behavior of $e_{-}^{\infty}$ and $e_{+}^{\infty}$, where $r$ is the dominant reconnection point also called X-point. 
In Fig. {\bf \ref{fig2}} we can see that $e^{\infty}_{-}$ is almost always less than $0$, while $e^{\infty}_{+}$ is always greater than $0$. Another fundamental requirement for energy extraction is that $e^{\infty}_{-}$ must be negative. Therefore, the condition $e^{\infty}_{-}<0$ is essential for the energy extraction process. In Fig. {\bf\ref{fig2}} we can see that, with the increase in $\sigma_0$, $e^{\infty}_{-}$ decrease and $e^{\infty}_{+}$  increases. In the left panel of Fig. {\bf\ref{fig2}}, as azimuthal angle $\xi$ increase $e^{\infty}_{-}$ increase while $e^{\infty}_{+}$ decreases. In the second panel of Fig. {\bf\ref{fig2}} we can see that as $a$ increase, $e^{\infty}_{-}$ decreases while $e^{\infty}_{+}$ increase.
In Fig. {\bf\ref{fig3}} we can see that, for different values of $\alpha$, neither $e^{\infty}_{-}$ nor $e^{\infty}_{+}$ varies monotonically. But the behavior of both $e^{\infty}_{-}$ and $e^{\infty}_{+}$ depends strongly on the location of dominant reconnection point (or X-point). Moreover, in both panels of Fig. {\bf\ref{fig3}}, for small values of $r$, the differences between dashed lines corresponding to the different values of $\alpha$ become more pronounced and clearly distinguishable.

\subsection{Parameter Space for Energy Extraction Via Magnetic Reconnection in Circular Orbits}
In this section, we plot the allowed energy extraction region ($e^{\infty}_{-}<0$ ) in the $r-a$ plane. In all these results, the blue solid line represents the ergosphere, the red solid line represents the event horizon, the purple dashed line represents the photon sphere radius, and from left to right $\sigma_0=100,~30,~10,~3$. 
\begin{figure}[H]
\begin{center}
\subfigure[~$\alpha=0.01$,~$\xi=\pi/12$]{\includegraphics[width=4.7cm,height=4.3cm]{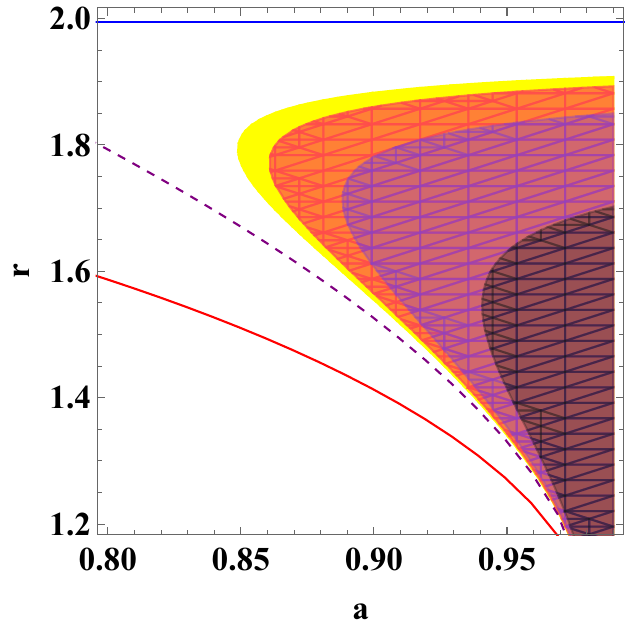}}
\subfigure[~$\alpha=0.1$,~$\xi=\pi/12$]{\includegraphics[width=4.7cm,height=4.3cm]{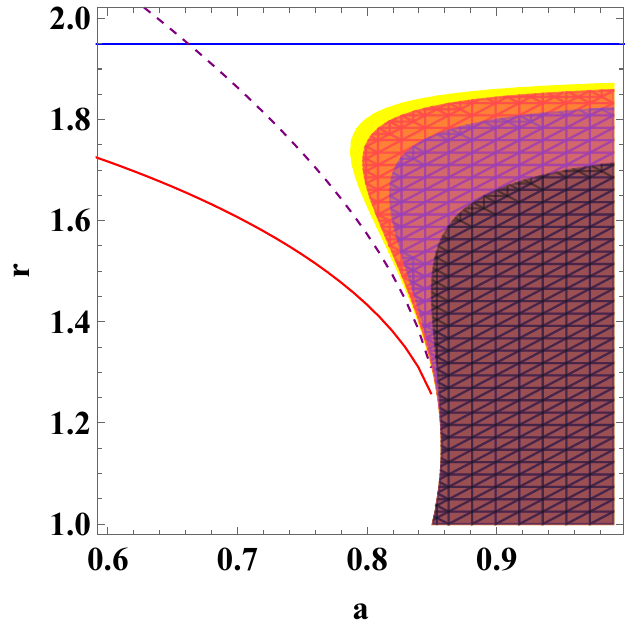}}
\subfigure[~$\alpha=0.2$,~$\xi=\pi/12$]{\includegraphics[width=4.7cm,height=4.3cm]{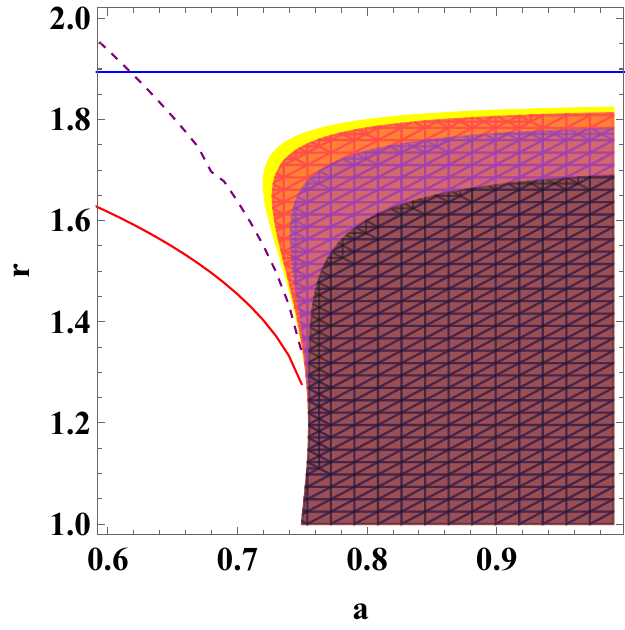}}
\caption{Plots showing the allowed energy extraction regions for different values of $\alpha$.}
\label{fig4}
\end{center}
\end{figure}
As shown in Fig. {\bf\ref{fig4}}, the energy extraction region expands as the value of $\sigma_0$ increases. This behavior is consistent with the results reported in Ref. \cite{comisso2021magnetic}. In Fig. {\bf\ref{fig4}}, we can see that the maximum allowed spin remain unchanged while the minimum allowed spin for energy extraction decreases  from $0.8$ in Fig. {\bf\ref{fig4}} (a) to $0.7$ in Fig. {\bf\ref{fig4}} (c), with the increasing value of $\alpha$. Moreover, due to the increase in event horizon, photon sphere radius and decrease in ergosphere, the position of the reconnection increases.
\begin{figure}[H]
\begin{center}
\subfigure[~$\alpha=0.1,~\xi=\pi/12$]{\includegraphics[width=4.7cm,height=4.3cm]{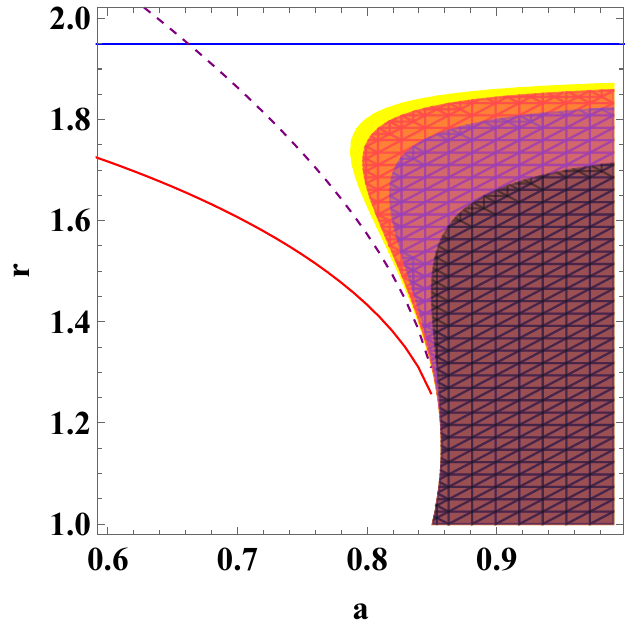}}
\subfigure[~$\alpha=0.1,~\xi=\pi/6$]{\includegraphics[width=4.7cm,height=4.3cm]{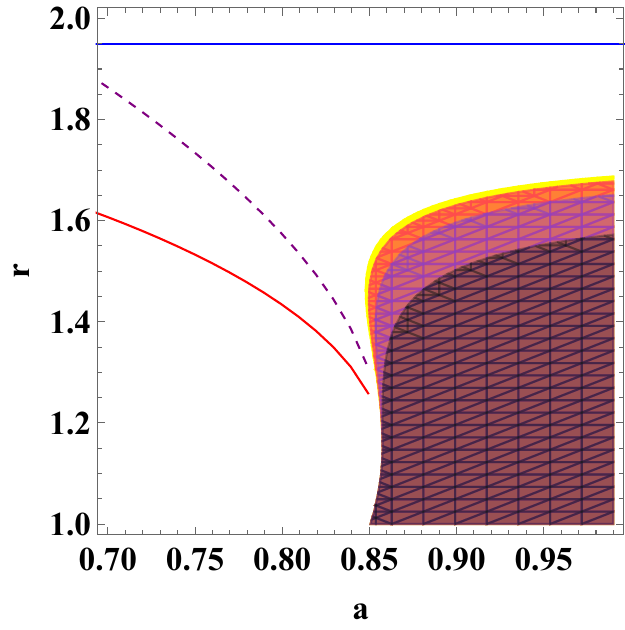}}
\subfigure[~$\alpha=0.1,~\xi=0$]{\includegraphics[width=4.7cm,height=4.3cm]{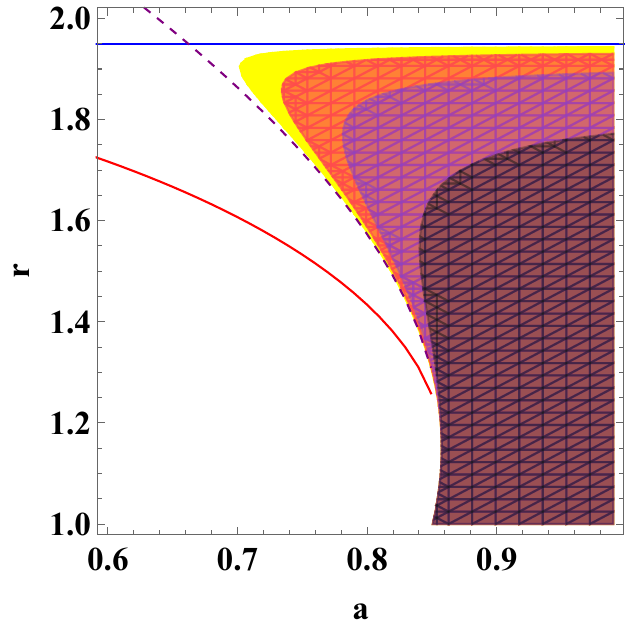}}
\caption{Plots showing the allowed energy extraction regions for different values of $\xi$.}
\label{fig5}
\end{center}
\end{figure}
In Fig. {\bf\ref{fig5}}, we can see that the area of the allowed region for energy extraction increases with decreasing the value of $\xi$, and the minimum allowed spin energy extraction decreases. The position of the reconnection layer more pronounced. The reconnection process shows no significant dependence on the value of $\xi$. Therefore, for convenience, we set $\xi=\pi/12$ in the subsequent analysis.
\begin{figure}[H]
\centering 
\includegraphics[width=0.5\linewidth]{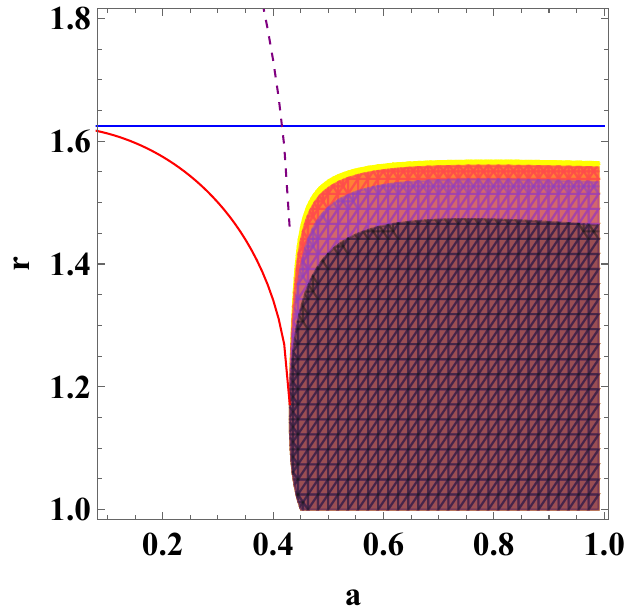}
\caption{Plot showing the allowed energy extraction region with a fixed value of $\alpha=0.61$.}
\label{fig5extra}
\end{figure}
As shown in Fig. {\bf\ref{fig5extra}}, there exists a minimum value of the spin parameter required for magnetic reconnection to occur. We reveal that for the larger value of $\alpha$, there is low spin. Furthermore, our results show that the allowed spin for energy extraction lies between $0.4$ and $0.98$, demonstrating that the energy extraction process can take place even at low BH spins.

\subsection{Power and Efficiency of Energy Extraction in Circular Orbits}
\begin{figure}[H]
\begin{center}
\subfigure[~$a=0.97,~\xi=\pi/12,~\alpha=0.01$]{\includegraphics[width=8cm,height=5cm]{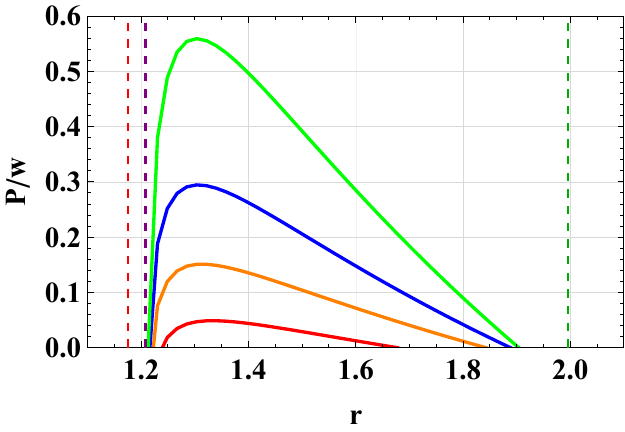}}
\subfigure[~$a=0.95,~\xi=\pi/12,~\alpha=0.01$]{\includegraphics[width=8cm,height=5cm]{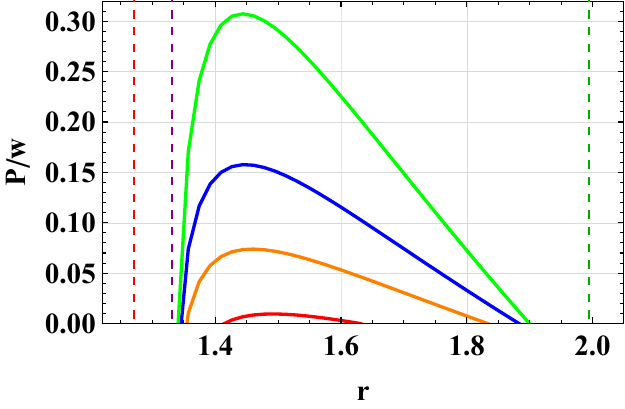}}
\caption{Plots showing the energy extraction power for different
values of $\sigma_0$.}
\label{fig7}
\end{center}
\end{figure}

\begin{figure}[H]
\begin{center}
\subfigure[~$a=0.852,~\xi=\pi/12,~\alpha=0.1$]{\includegraphics[width=8cm,height=5cm]{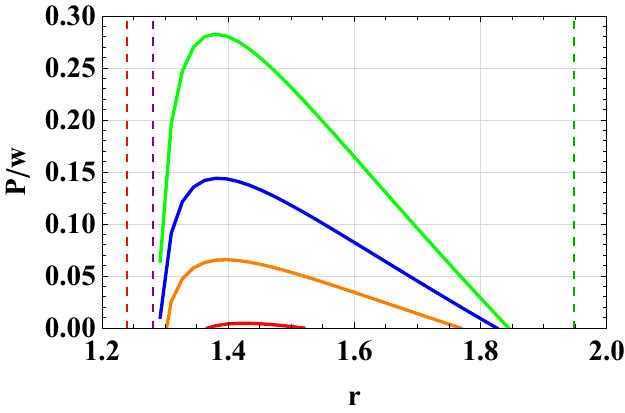}}
\subfigure[~$a=0.755,~\xi=\pi/12,~\alpha=0.2$]{\includegraphics[width=8cm,height=5cm]{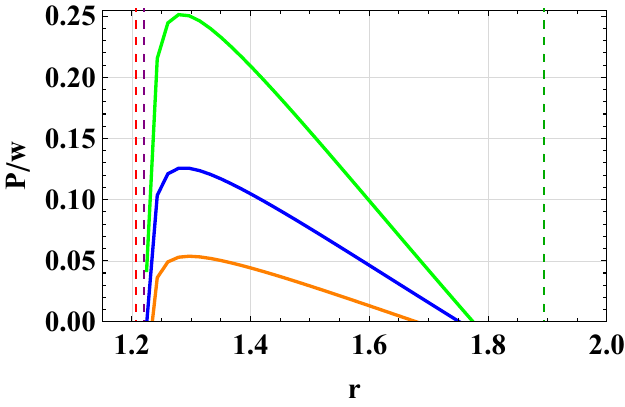}}
\caption{Plots showing the energy extraction power for different
values of $\sigma_0$.}
\label{fig8}
\end{center}
\end{figure}

After having the feasibility of energy extraction, now we compare the power and efficiency. So, the energy extraction power is given as \cite{comisso2021magnetic}
\begin{equation}
P=-e_{-}^{\infty}\omega A_{in}U_{in},
\label{power}
\end{equation}
where, $U_{in}$ is used for two conditions: one for collisionless conditions, $U_{in}\approx 0.1$ \cite{comisso2016value} and the other for collisional conditions, $U_{in}\approx 0.01$ \cite{huang2010scaling}. We use a collisionless condition $U_{in}\approx 0.1$ in this paper. Moreover, $A_{in}$ is the cross-sectional area of the inflowing plasma and can be expressed as
\begin{equation}
A_{\rm in} \sim (r_{E}^{2}-r_{p}^{2}).
\label{area}
\end{equation}
where $r_{E}$ is the boundary of ergosphere and $r_{P}$ is the photon sphere radius. In Figs. {\bf\ref{fig7}} and {\bf\ref{fig8}}, we plot the energy extraction power per enthalpy density, $P/\omega$ with respect to $r$. In these figures, green, blue, orange and red solid curves represent to $\sigma_0=100,~30,~10$ and $3$, respectively. And the purple, green and red dashed lines corresponds to the photon sphere, the boundaries of the ergosphere, and the event horizon, respectively. 
From the left panel of Fig. {\bf\ref{fig7}}, we can see that the energy extraction power start from the outside of the photon sphere radius, in agreement with the characteristics of circular orbits. We also observe that the energy extraction power initially increases with increasing $r$, reaches a maximum value, and then gradually decreases. We also find that the energy extraction power increases with increasing $\sigma_0$. By comparing Figs. {\bf\ref{fig7}} (a) and {\bf\ref{fig8}} (b), we observe that, for the same value of $\sigma_0$, the energy extraction power decreases as the GB coupling parameter $\alpha$ increases. It should also be noted that the spin parameter $a$ is not chosen to be the same in both cases, since the BH spin depends on $\alpha$. A similar behavior was reported in Ref. \cite{li2023energy}, where the allowed range of the spin parameter varies for different values of $h_0$.  Also, by comparing Fig. \ref{fig7}(b) and \ref{fig8}(b), we can see that the energy extraction power decreases at higher values of $\alpha$. In this case also the spin parameter $a$ cannot be remain same.
Now, we plot the energy extraction efficiency. The energy extraction efficiency is given as \cite{comisso2021magnetic} 
\begin{equation}
\eta=\frac{e_{+}^{\infty}}{e_{+}^{\infty}+e_{-}^{\infty}}.
\label{efficiency}
\end{equation}
Note that since energy extraction requires $e_{-}^{\infty}<0$ and $e_{+}^{\infty}>0$, the efficiency defined by the above expression is meaningful only when $\eta>1$. For this we fix $\xi=\pi/12$ and $\sigma_0=100$.
\begin{figure}[H]
\begin{center}
\subfigure[~$\alpha=0.01$]{\includegraphics[width=8cm,height=5cm]{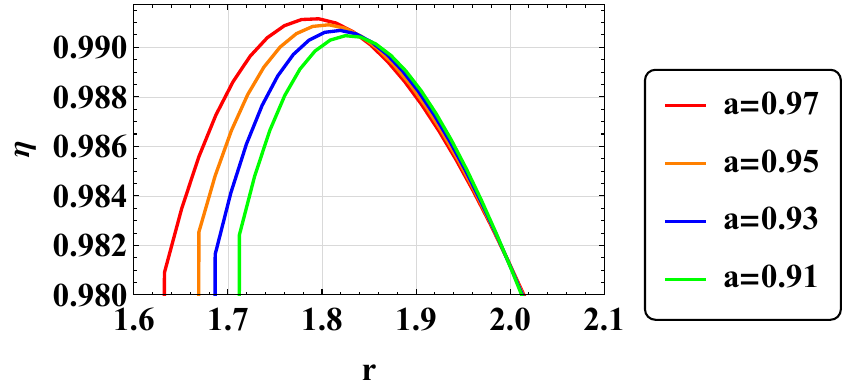}}
\subfigure[~$ \alpha=0.01$]{\includegraphics[width=8cm,height=5cm]{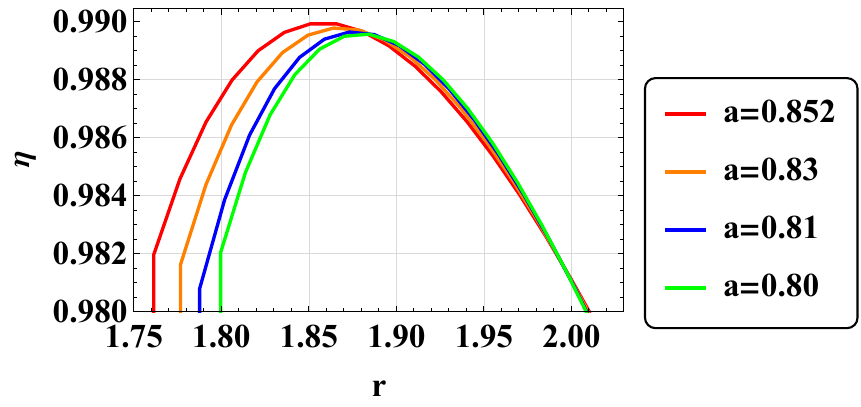}}
\caption{Plots showing the energy extraction efficiency for different
values of $a$.}
\label{fig9}
\end{center}
\end{figure}
In Fig. {\bf\ref{fig9}} (a), we plot the energy extraction efficiency under the different values of the spin parameter $a$. As in the same case of energy extraction power, energy extraction efficiency also increases with $r$, reaches a peak value, and then decreases gradually. Also, we can observe that at the higher value of $a$, the energy extraction efficiency is higher. By comparing Figs. {\bf\ref{fig9}} (a) and {\bf\ref{fig9}} (b), it should be noted that the values of $a$ are not same because the allowed spin changes under the different values of $\alpha$. A comparison of Figs.  {\bf\ref{fig10}} (a) and {\bf\ref{fig10}} (b) also shows that the energy extraction power increases with increasing the value of $\alpha$, which is similar to the power trend. If we compare all the panels of Figs. \ref{fig9} and \ref{fig10}, it can be noticed that the energy extraction efficiency is higher at the lower value of $a$, which also follows the power trend. 
\begin{figure}[H]
\begin{center}
\subfigure[~$\alpha=0.1$]{\includegraphics[width=8cm,height=5cm]{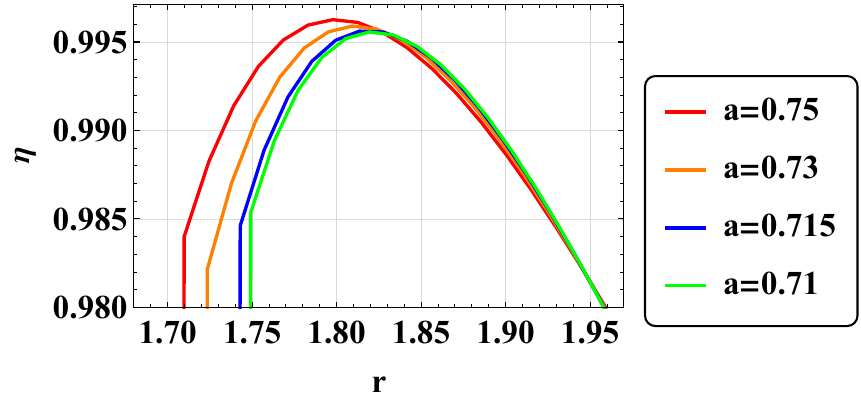}}
\subfigure[~$\alpha=0.2$]{\includegraphics[width=8cm,height=5cm]{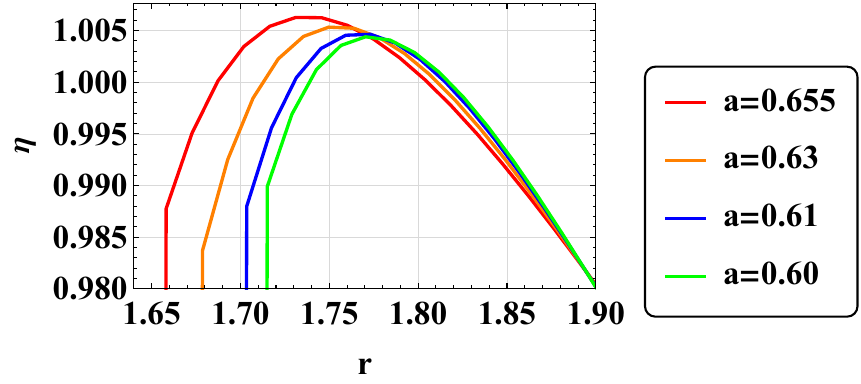}}
\caption{Plots showing the energy extraction efficiency for different
values of $a$.}
\label{fig10}
\end{center}
\end{figure}
Now we compare the energy extraction power ratio with that of the
BZ framework \cite{blandford1977electromagnetic}. The
corresponding BZ energy extraction power can be
defined as \cite{tchekhovskoy2010black,camilloni2022blandford}:
\begin{equation}
P_{\text{BZ}}=\frac{\kappa}{16\pi} \, \Phi_H^2 \big(
\Omega_H^2+\alpha_{1}\Omega_H^4
+\alpha_{2}\Omega_H^6\big),\label{blandfordpowe}
\end{equation}
in which $\kappa=0.05$, $\alpha_1=1.38$ and $\alpha_2=-9.2$ are
numerical constants, and $\Omega_H$ correspond to the angular
velocity of the event horizon, which is expressed as
\begin{equation}
\Omega_H = \left. \frac{-g_{t\phi}}{g_{\phi\phi}} \right|_{r=r_+}=
\frac{a}{r_+^2  + a^2}. \label{eventhorizonvelocity}
\end{equation}
In Eq. (\ref{blandfordpowe}) $\Phi_H$ represent the magnetic flux
threading one hemisphere of the BH event horizon, which has
the following form as
\begin{equation}
\Phi_H = \frac{1}{2} \iint |B^r| \sqrt{g_{\theta\theta}
g_{\phi\phi}} \, d\theta d\phi = 2\pi \left( r_+^2  + a^2
\right)\ \, B_0 \sin\xi, \label{magneticflux}
\end{equation}
where $B_0={(w\sigma_{0}})^{1/2}$. Now, we calculate the power ratio by
following a straight forward mechanism. Specifically, by squaring the
above expression, yields
\begin{equation}
\Phi_H^2 = 4\pi^2 \left( r_+^2  + a^2 \right)^2
\, B_0^2 \sin^2\xi,
\label{squaringflux}
\end{equation}
Substituting $\Phi_H^2$ in Eq. (\ref{blandfordpowe}), we obtain
\begin{align}
P_{\text{BZ}} &=\frac{\kappa}{16\pi} \left(4\pi^2 \left(
r_+^2  + a^2 \right)^2 \right)  B_0^2 \sin^2\xi \left(
\Omega_H^2+ \alpha_1 \Omega_H^4 + \alpha_2  \Omega_H^6 \right).
\label{Bzfluxsubstitute}
\end{align}
After simplifying, we have
\begin{equation}
P_{\text{BZ}}=\frac{\kappa \pi}{4} \left( r_+^2  + a^2\right)^2 \, B_0^2 \sin^2\xi \left( \Omega_H^2+ \alpha_1\Omega_H^4 + \alpha_2 \Omega_H^6 \right). \label{constantputting}
\end{equation}
So, the reconnection power is given by $P=- e^{\infty}_{-} A_{\text{in}}
U_{\text{in}} \, \omega$. By dividing this expression by the power
$P_{\text{BZ}}$, we obtain the corresponding normalized power ratio
as follows
\begin{align}
\frac{P}{P_{\text{BZ}}} &= \frac{- e^{\infty}_{-} A_{\text{in}} U_{\text{in}}
\, \omega}{\frac{\kappa \pi}{4} \left( r_+^2 +  a^2
\right)^2 \, \omega \sigma_0 \sin^2\xi \left(\Omega_H^2+
\alpha_1 \Omega_H^4 + \alpha_2 \Omega_H^6 \right)},\label{PBSpower}
\end{align}
which leads that
\begin{equation}
\frac{P}{P_{\text{BZ}}}=\frac{-4 e^{\infty}_{-} A_{\text{in}} U_{\text{in}}} {\kappa \pi \sigma_{0} \left( \Omega_H^2+ \alpha_1
\Omega_H^4 + \alpha_2 \Omega_H^6 \right) \sin^2\xi \left( r_+^2+a^2
\right)^2}. \label{powerRatio}
\end{equation}
The Eq. (\ref{powerRatio}) shows that, for sufficiently small
values of the azimuthal angle, the power ratio can exceed unity.
This suggests that, under such conditions, the magnetic reconnection
mechanism can yield a higher energy extraction power than the BZ framework. So, for conveniently we fix
$\xi=\pi/12$. For the power comparison, we pick a set of parameters for high spin ($a=0.98,~r=1.1,~\sigma_0=5,~\alpha=0.01$), and we gain a power ratio $11.5537>1$, which show that magnetic reconnection power is higher than BZ power. We also pick a set of parameters for low spin $a=0.5,~r=2.3,~\sigma_0=10,~\alpha=0.25$, and we get the power ratio $5.06476$, which indicates that even though the absolute power and efficiency are reduced, the magnetic reconnection power also exceeds the BZ power. This behavior is attributed to the influence of the GB coupling parameter $\alpha$, which lowers the spin threshold and enhances the efficiency of magnetic reconnection.

\section{EXTRACTING four-dimensional EGB BH ENERGY IN THE PLUNGING REGION}

\subsection{Magnetic Reconnection Process in the Plunging Region}

\begin{figure}[H]
\begin{center}
\subfigure[~$\alpha=0.01$,~$\xi=\pi/12$]{\includegraphics[width=4.7cm,height=4.3cm]{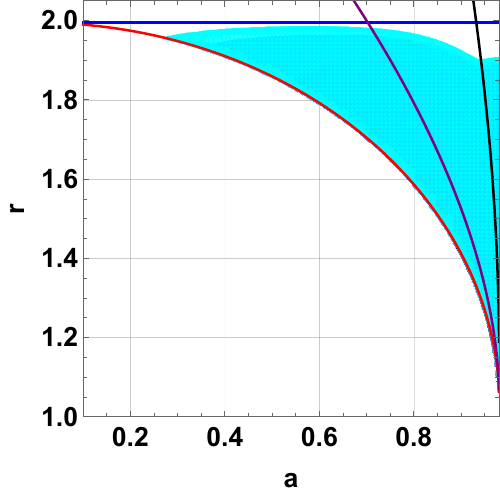}}
\subfigure[~$\alpha=0.1$,~$\xi=\pi/12$]{\includegraphics[width=4.7cm,height=4.3cm]{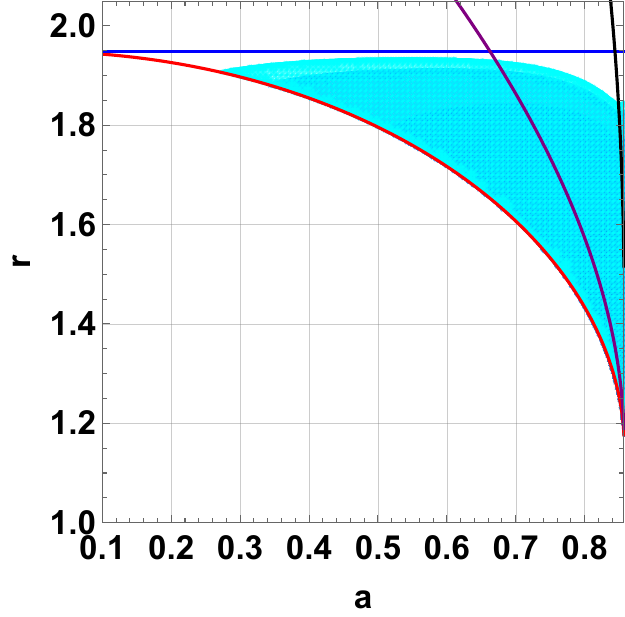}}
\subfigure[~$\alpha=0.2$,~$\xi=\pi/12$]{\includegraphics[width=4.7cm,height=4.3cm]{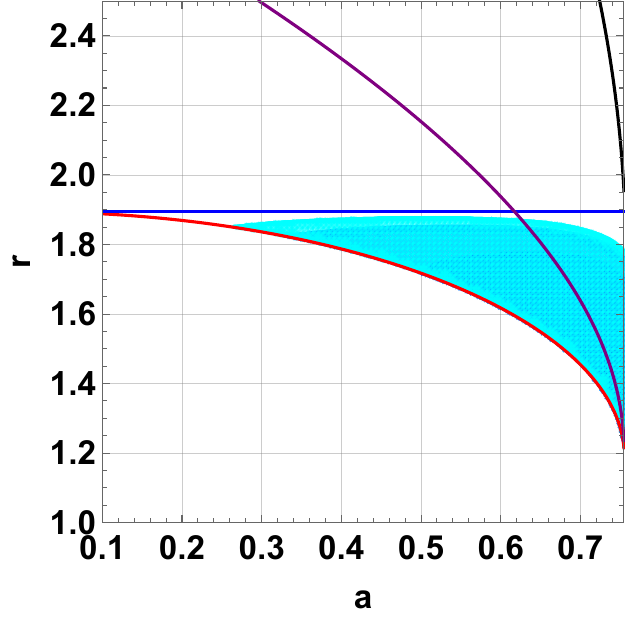}}
\caption{Plots showing the allowed energy extraction regions for different values of $\alpha$ in the plunging region.}
\label{fig11}
\end{center}
\end{figure}

\begin{figure}[H]
\begin{center}
\subfigure[~$\alpha=0.1,~\xi=\pi/12$]{\includegraphics[width=4.7cm,height=4.3cm]{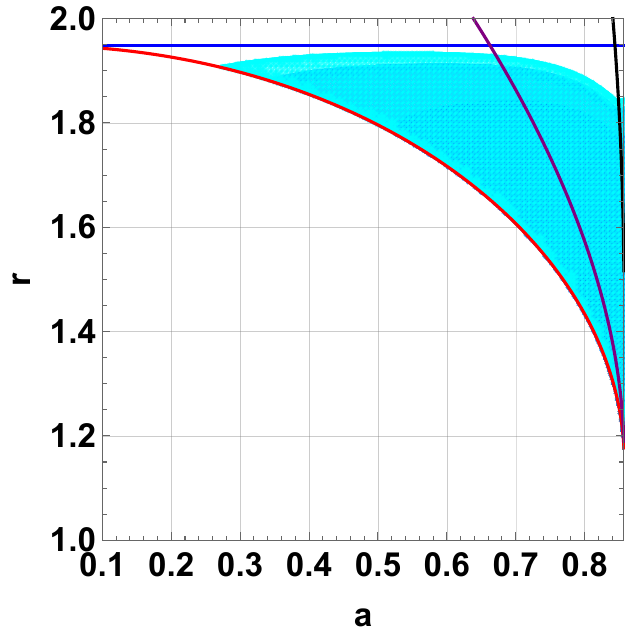}}
\subfigure[~$\alpha=0.1,~\xi=\pi/6$]{\includegraphics[width=4.7cm,height=4.3cm]{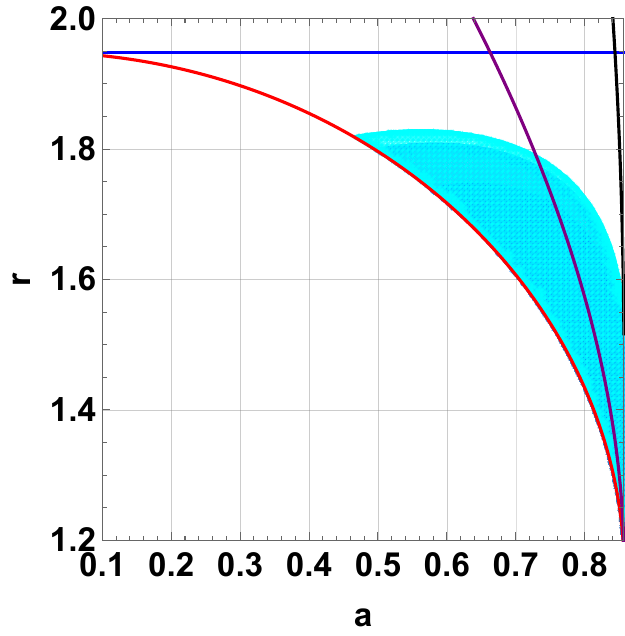}}
\subfigure[~$\alpha=0.1,~\xi=0$]{\includegraphics[width=4.7cm,height=4.3cm]{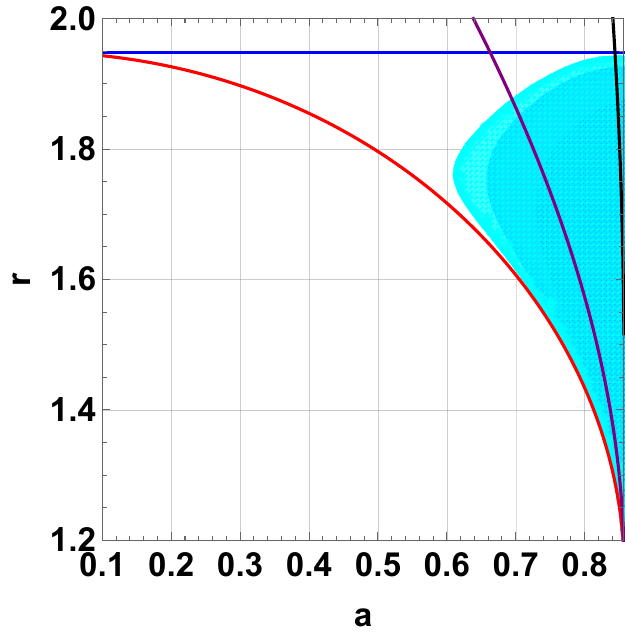}}
\caption{Plots showing the allowed energy extraction regions for different values of $\xi$ in the plunging region.}
\label{fig12}
\end{center}
\end{figure}

\begin{figure}[H]
\centering
\includegraphics[width=0.3\linewidth]{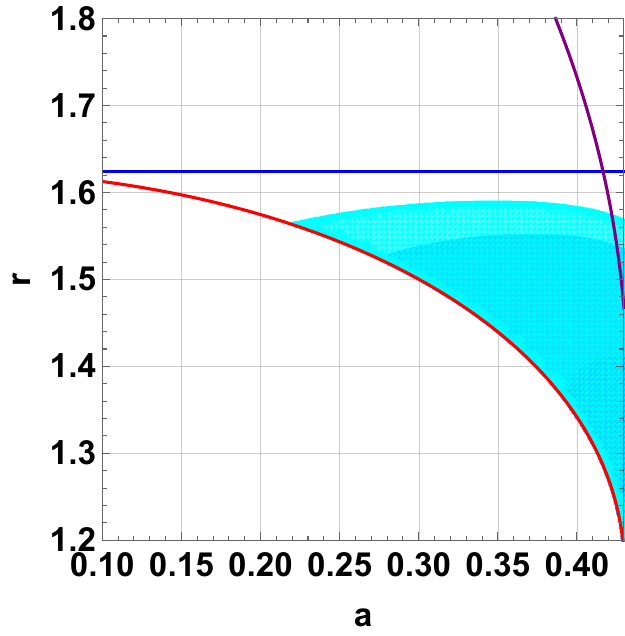}
\caption{Plot showing the allowed energy extraction regions for a fixed $\alpha=0.61$ and $\xi=\pi/12$ in the plunging region.}
\label{fig13}
\end{figure}

We then consider the plunging region and examine the energy extraction process within this regime. Initially, the plasma follows a circular orbit just outside the ISCO. However, circular orbits become unstable inside the ISCO radius. As a result, any perturbation initiates radial motion, causing the plasma to plunge inward from the ISCO. This inner region is commonly referred to as the plunging region. Here, the presence of the significant radial velocity invalidates the Keplerian velocity formula given in (\ref{epsilon}), as that expression accounts solely for the azimuthal component. For a detailed analytical and numerical discussion of the plunging region, see Ref. \cite{chen2024energy}. In this section, we briefly discuss the energy extraction process in the plunging region. The transformation of four-velocity components between the BL coordinates and the ZAMO frame is expressed as 
\begin{equation}
U^{\mu}=\hat{\gamma}_{s}\begin{pmatrix}1 \\\hat{v}_{s}^{(r)} \\0 \\\hat{v}_{s}^{(\phi)}\end{pmatrix}=\begin{pmatrix}\dfrac{E-\omega^{\phi}L}{N} \\\sqrt{g_{rr}}\,U^{r} \\0 \\ \dfrac{L}{\sqrt{g_{\phi\phi}}} \end{pmatrix},
\label{BLcordinates}
\end{equation}
where
\begin{equation}
U^{r}=\frac{dr}{d\tau},\qquad\left(U^{r}\right)^{2}=\left(\frac{dr}{d\tau}\right)^{2},
\label{urequation}
\end{equation}
where $(dr/d\tau)^2$ satisfies (\ref{radialfunction}). In the plunging region, the energy $E$ and angular momentum $L$ are conserved quantities and their values at ISCO are replaced by 
\begin{equation}
L_{I}=L(r_{I}),\qquad E_{I}=E(r_{I})
\label{replacement}
\end{equation}
where $r_{I}$ is the radius of ISCO, and $E$ and $L$ satisfy (\ref{expressionforEandL}). 
For ISCO, the conditions are 
\begin{equation}
R(r)=0,\qquad R'(r)=0,\qquad R''(r)=0.
\label{conditionforISCO}
\end{equation}
Substituting (\ref{replacement}) into (\ref{radialfunction}), and using $\varepsilon=1$, we get
\begin{align}
U^r
= -\frac{1}{\Sigma^2}\sqrt{[(r^{2}+a^{2})E_{I} - aL_{I}]^2 - \Delta_r\left[(L_{I}-aE_{I})^2 + \mathcal{O}\right]}.
\label{uraftersubstution}
\end{align}
The negative sign in front of the above equation indicates the inward motion.
Substituting (\ref{uraftersubstution}) into (\ref{urequation}) gives
$\hat{v}_{s}^{(r)}$,
$\hat{v}_{s}^{(\phi)}$,
and
\begin{equation}
\hat{v}_{s}=\sqrt{\left(\hat{v}_{s}^{(r)}\right)^{2}+\left(\hat{v}_{s}^{(\phi)}\right)^{2}},
\label{velocityv}
\end{equation}
$\hat{\gamma}_{s}$ is the Lorentz factor of $\hat{v}_{s}$,
$r_I$ satisfies (\ref{conditionforISCO}).
In this case,
$e_{\pm}^{\infty}$ becomes as \cite{comisso2021magnetic,chen2024energy}
\begin{equation}
\begin{aligned}
e_{\pm}^{\infty}=&\,N \hat{\gamma}_{s}\gamma_{\rm out}\Bigg[\left(1+\beta^{\phi}\hat{v}_{s}^{(\phi)}\right)\pm v_{\rm out}\left(\hat{v}_{s}+\beta^{\phi}\frac{\hat{v}_{s}^{(\phi)}}{\hat{v}_{s}}\right)\cos\xi\mp v_{\rm out}\beta^{\phi}\frac{\hat{v}_{s}^{(r)}}{\hat{\gamma}_{s}\hat{v}_{s}}\sin\xi\Bigg]\\&-N\left[4\hat{\gamma}_{s}\gamma_{\rm out}\left(1\pm \hat{v}_{s}v_{\rm out}\cos\xi\right)\right]^{-1},
\end{aligned}
\label{epsiloninplunging}
\end{equation}
where $v_{\rm out}$ represents the outflow velocity and $\gamma_{\rm out}$ denotes its corresponding Lorentz factor. These quantities are defined by the magnetization parameter $\sigma_{0}$ as
\begin{equation}
v_{\rm out}=\sqrt{\frac{\sigma_{0}}{\sigma_{0}+1}},\qquad\gamma_{\rm out}=\sqrt{1+\sigma_{0}}.
\label{lorentzplunging}
\end{equation}
As in the circular orbit case, we now plot the energy extraction region in the $r-a$ plane for the plunging region, as shown in Figs. {\bf\ref{fig11}} and {\bf\ref{fig12}}. In these figures, the red, blue, purple, and black curves represent the event horizon, ergosphere, photon sphere radius, and ISCO, respectively. From left to right, the panels correspond to $\sigma_0=100,~30,~10,~3$.
From Fig. {\bf\ref{fig11}} (a), we can see that when $\sigma_0$ increases, the allowed energy extraction region also increases. Comparing Figs. {\bf\ref{fig11}} (a) and {\bf\ref{fig4}} (a), we find that for $r>r_I$, the allowed energy extraction region closely matches that of the circular orbit case. In contrast, for $r < r_I$, the allowed region in the plunging regime is significantly larger than that in the circular orbit case. We also observe that the minimum allowed spin is lower in the plunging region than in the circular orbit case. In addition, the position of the reconnection point is shifted to larger radii in the plunging region. For example, for $\sigma_{0}=100$, the minimum allowed spin is $0.86$ in the circular orbit case Fig. {\bf\ref{fig4}} (a), whereas it decreases to $0.28$ in the plunging region. A comparison of the three panels in Fig. {\bf\ref{fig11}} further shows that the minimum allowed spin remains constant at $0.28$ with increasing values of $\alpha$, but the maximum allowed spin decreases from $0.86$ in Fig.  {\bf\ref{fig11}} (a) to $0.76$ in Fig. {\bf\ref{fig11}} (c). And the position of the boundary of the ergosphere also decreases as $\alpha$ increases. This indicate that energy can be extracted from slowly rotating BHs. From Fig. {\bf\ref{fig12}}, we observe that the minimum allowed spin increases with decreasing azimuthal angle $\xi$. We also observe that the shape of the allowed energy extraction region changes with the azimuthal angle. This dependence is a characteristic feature of the plunging region.
Following Fig. {\bf\ref{fig5extra}}, we plot the allowed region for large values of $\alpha$ in Fig. {\bf\ref{fig13}}. Here we can see that the minimum allowed spin for energy extraction is $0.22$.

\subsection{Power of Energy Extraction in the Plunging Region}

In this section, we plot the energy extraction power for the plunging region and circular orbits. After comparison, we indicate the difference between the plunging case and the circular orbit case in the region where the condition $r_I>r$ is satisfied. We pick the set of parameters like $a=0.90,~\xi=\pi/12,~\sigma_0=100$ and $\alpha=0.01$. In this case all values of ISCO are greater than $r$.
\begin{figure}[H]
\centering  
\includegraphics[width=0.5\linewidth]{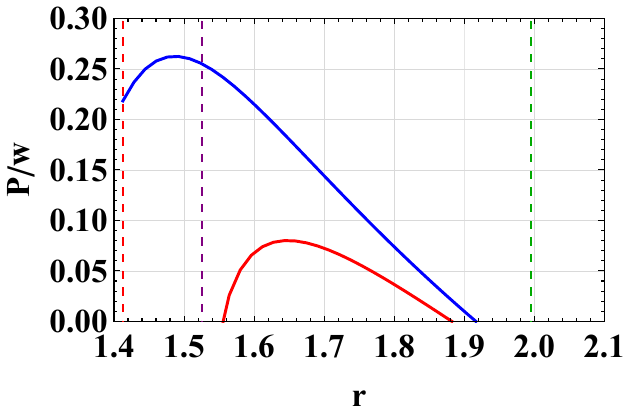}
\caption{Red solid and blue solid lines represent the circular orbit power and plunging region power, respectively. Purple dashed, red dashed and green dashed lines correspond to the photon sphere radius, event horizon, and ergosphere, respectively.}
\label{fig14}
\end{figure}
From Fig. {\bf\ref{fig14}}, we can see that the power of energy extraction in the plunging region is higher than the power in a circular orbit. Finally, we compare the energy extraction power in the plunging region with that of the BZ process using Eq. (\ref{powerRatio}). For this comparison, we use the same parameter values as those adopted in the circular orbit case, i.e., $a=0.98,~r=1.1,~\sigma_0=5,~\alpha=0.01$. After putting these values, we can notice that $r_I=1.24665>r$, which satisfies the plunging region condition. We gain a power ratio of $28.8724$, which is greater than $11.5537$, obtained in a circular orbit. This result again indicates that the plunging region power is higher than both the circular orbit power and BZ power. Now considering a low spin case based on Fig. {\bf\ref{fig11}} (a) where the set of parameters is $\alpha=0.001,~\xi=\pi/12$, and spin $a$ can be $0.3$ or lower. Here we note that the ergosphere is below the photon sphere radius, so Eq. (\ref{area}) is no longer valid and should be replaced by 
\begin{equation}
A_{in}\sim(r^2_{E}-r^2_{+}).
\label{new area}
\end{equation}
If we take $\sigma_0=100,~a=0.3,~r=4.3$ and $\xi=\pi/12$, then we gain a power ratio of $0.0859538$ which is very low. This is acceptable because a larger value of $\sigma_0$ is required to achieve low spin in the plunging region. The main reason is that, in Eq. (\ref{powerRatio}) $\sigma_0$ appears in the denominator, which reduces the power ratio. Nevertheless, it is remarkable that energy extraction through magnetic reconnection remains possible even for a spin parameter as low as $0.22$, which is lower than the previously reported threshold.

\section{CONCLUSION}

In this work, we have investigated the extraction of rotational energy from a rotating BH in four-dimensional EGB gravity through the magnetic reconnection mechanism. We first examined the geometrical properties of the spacetime, including the event horizon, ergosphere, and photon sphere, and analyzed how these quantities are modified by the GB coupling parameter. We found that increasing the coupling parameter changes the size of the ergosphere and reduces the maximum allowed BH spin, thereby affecting the conditions for energy extraction.
\par For plasma in circular orbits, we studied the magnetic reconnection process by analyzing the energy at infinity of the accelerated and decelerated plasma outflows. The allowed parameter space demonstrates that the magnetic reconnection mechanism remains effective over a broad range of parameters and, importantly, the inclusion of the GB coupling lowers the minimum spin required for energy extraction. Our results show that energy extraction is possible for BH with spin as low as $a \simeq 0.4$, significantly below the threshold reported for the Kerr case. We also investigated the corresponding energy extraction power and efficiency, and we observed that both quantities strongly depend on the plasma magnetization and the location of the reconnection layer. Furthermore, the normalized power comparison shows that the magnetic reconnection mechanism can produce a larger energy extraction power than the BZ process, not only for rapidly rotating BHs but also for moderately rotating ones due to the influence of the GB coupling.
\par We further extended the analysis to the plunging region, where the plasma departs from stable circular motion after crossing the ISCO. We found that the allowed energy extraction region becomes substantially larger than that of the circular orbit case, particularly inside the ISCO, and that energy extraction remains possible for even lower BH spins as $0.22$. The GB coupling continues to reduce the required spin threshold, while the dependence of the allowed region on the magnetic field orientation becomes more pronounced than in circular orbits. Moreover, the energy extraction power in the plunging regime is found to exceed that of the circular orbit regime, indicating that plasma plunging toward the event horizon provides a more favourable environment for magnetic reconnection and rotational energy extraction.
\par Overall, our results demonstrate that higher curvature corrections encoded in four-dimensional EGB gravity significantly enhance the effectiveness of magnetic reconnection as an energy extraction mechanism. The reduction of the spin threshold together with the enhanced power output suggests that magnetic reconnection may operate efficiently even around slowly rotating BHs in modified gravity. These findings provide new insights into high-energy processes occurring in BH magnetospheres and may offer a useful theoretical framework for testing deviations from GR through future observations of accreting BHs and relativistic jets.

\end{document}